\documentclass[aps,prl,twocolumn]{revtex4-1}
\usepackage{amsfonts, amssymb, amsmath,graphicx}
\usepackage{subfigure}
\usepackage{multirow}
\usepackage{dcolumn}
\usepackage{bm}
\usepackage{color}

\begin{document}

\title{Steady-state Hall response and quantum geometry of driven-dissipative lattices}
\author{Tomoki Ozawa}
\affiliation{INO-CNR BEC Center and Dipartimento di Fisica, Universit\`a di Trento, I-38123 Povo, Italy}%

\date{\today}

\newcommand{\iac}[1]{{\color{red} #1}}
\newcommand{\tom}[1]{{\color{blue} #1}}

\makeatletter
\def\@tvsp{\mathchoice{{}\mkern-4.5mu}{{}\mkern-4.5mu}{{}\mkern-2.5mu}{}}
\def\llangle{\langle\@tvsp\langle}
\def\rrangle{\rangle\@tvsp\rangle}
\makeatother

\begin{abstract}
We study the effects of the quantum geometric tensor, i.e., the Berry curvature and the Fubini-Study metric, on the steady state of driven-dissipative bosonic lattices.
We show that the quantum-Hall-type response of the steady-state wave function in the presence of an external potential gradient depends on all the components of the quantum geometric tensor.
Looking at this steady-state Hall response, one can map out the full quantum geometric tensor of a sufficiently flat band in momentum space using a driving field localized in momentum space.
We use the two-dimensional Lieb lattice as an example and numerically demonstrate how to measure the quantum geometric tensor.
\end{abstract}

\maketitle

The developments in the study of topological insulators and superconductors have led to growing interest in understanding the geometry and topology of energy bands in condensed-matter physics~\cite{Xiao:2010, Hasan:2010, Qi:2011}. The prototype of the two-dimensional topological insulator is the quantum Hall system, where the quantized Hall conductance is related to the topological Chern number of bands~\cite{Klitzing:1980, Thouless:1982}. The Chern number is the integral of the geometrical Berry curvature over the Brillouin zone. This Berry curvature is associated with the adiabatic response of the system under a slow change in parameters~\cite{Berry:1984} and has been directly measured in experiments~\cite{Flaschner:2016, Wimmer:2017}.

Along with these developments, another geometrical quantity of energy bands, the Fubini-Study metric, has recently attracted great interest~\cite{Kolodrubetz:2016, Provost:1980, Anandan:1990, Albert:2016, Ma:2010, Legner:2013, Roy:2014, Claassen:2015, Bauer:2016, Julku:2016, Liang:2017, CamposVenuti:2007, Zanardi:2007, You:2007, Dey:2012, Kolobrubetz:2013, Neupert:2013, Srivastava:2015, Lim:2015PRA, Bleu:2016, Raoux:2015, Gao:2015, Piechon:2016, Combes:2016, Freimuth:2017}.
The Fubini-Study metric describes the ``distance" between two quantum states in a parameter space and is related to phases acquired under non adiabatic processes~\cite{Provost:1980, Anandan:1990}. Therefore, it affects physical observables under the fast change of external parameters, or when the system is dissipative~\cite{Kolodrubetz:2016, Albert:2016}. It is also known to play important roles in the orbital magnetic susceptibility~\cite{Raoux:2015, Gao:2015, Piechon:2016, Combes:2016, Freimuth:2017}, the entanglement and many-body properties of quantum systems~\cite{Ma:2010, Legner:2013, Roy:2014, Claassen:2015, Bauer:2016}, superfluid density~\cite{Julku:2016, Liang:2017}, and quantum information~\cite{CamposVenuti:2007, Zanardi:2007, You:2007, Dey:2012}. Both the Berry curvature and the Fubini-Study metric can be understood from a general framework in terms of the quantum geometric tensor, whose real part gives the Fubini-Study metric while the imaginary part gives the Berry curvature~\cite{Kolodrubetz:2016}.

In this paper, we find a simple relation between the quantum-Hall-type response in the steady state of driven-dissipative bosonic lattices and the full quantum geometric tensor. Our results are relevant to photonic or mechanical lattices with non trivial geometrical or topological band structures, which have lately been an intensive area of study~\cite{Lu:2014, Lu:2016, Susstrunk:2015, Huber:2016}. Previous studies have shown that analogs of the quantum Hall effects take place in certain lattices with loss, through the center-of-mass response of the steady-state when a lattice site is continuously driven~\cite{Ozawa:2014, Ozawa:2016, Salerno:2016}. Here we clarify the relation between the full quantum geometric tensor and the center of mass within linear response for general driven-dissipative lattices, which is valid in any spatial dimensions.

Our finding can, in turn, be used to map out the full quantum geometric tensor of sufficiently flat bands over the entire Brillouin zone. There have been proposals to directly observe the effects of the Fubini-Study metric in various non adiabatic systems~\cite{Kolobrubetz:2013}, such as through the current noise spectrum in electronic systems~\cite{Neupert:2013}, the Lamb-shift-like energy shift in excitons~\cite{Srivastava:2015}, the St\"uckelberg interferometry for ultracold atoms~\cite{Lim:2015PRA}, and non adiabatic evolution of wave packets in exciton-polariton microcavities~\cite{Bleu:2016}.
One can also, in principle, use the quantum-state tomography technique~\cite{Flaschner:2016, Li:2016Science, Tarnowski:2017PRL} to construct the Fubini-Study metric from the full information of Bloch states.
There is, however, no direct experiment to date reporting the measurement of the Fubini-Study metric. Therefore, it is of great interest to propose concrete procedures from which one can experimentally determine the Fubini-Study metric. Our method to map out the full quantum gemetric tensor over the Brillouin zone can be a versatile and unique tool in topological photonics and topological mechanics to explore the topology and geometry of energy bands. In the final part of the paper, we use the flat band of the two-dimensional Lieb lattice as an illustration to numerically demonstrate our principle.  

\textit{Quantum geometric tensor.}
We consider non interacting tight-binding models with $q$ lattice sites per unit cell. The model has $q$ energy bands; $\mathcal{E}_n (\mathbf{k})$ is the dispersion of the $n$th band as a function of momentum $\mathbf{k}$ whose corresponding Bloch state is $|n_\mathbf{k}\rangle$. The Berry curvature $\Omega_{ij}^n (\mathbf{k})$ is defined by
\begin{align}
	\Omega_{ij}^n (\mathbf{k}) &\equiv i \left[ \langle \partial_{k_i} n_\mathbf{k} | \partial_{k_j}n_\mathbf{k} \rangle - \langle \partial_{k_j} n_\mathbf{k} | \partial_{k_i}n_\mathbf{k} \rangle \right],
	\label{berrydef}
\end{align}
which has a simple physical meaning as the magnetic field in momentum space~\cite{Adams:1959, Fang:2003, Bliokh:2005, Price:2014}.
The band has another geometrical structure called the Fubini-Study metric, which is defined through the following relation determining the ``distance" $ds^2$ between two nearby Bloch states
\begin{align}
	\hspace{-0.1cm}ds^2 = 1 - |\langle n_\mathbf{k}| n_{\mathbf{k} + d\mathbf{k}}\rangle|^2
	\equiv \sum_{i, j} g_{ij}^n dk_i dk_j + \mathcal{O}(|d\mathbf{k}|^3),
\end{align}
which leads to
\begin{align}
	&g_{ij}^n (\mathbf{k}) \equiv \frac{1}{2}
	\left[
	\langle \partial_{k_i} n_\mathbf{k} | \partial_{k_j}n_\mathbf{k} \rangle + \langle \partial_{k_j} n_\mathbf{k} | \partial_{k_i}n_\mathbf{k} \rangle 
	\right.
	\notag \\
	&\left.
	-\langle \partial_{k_i} n_\mathbf{k} | n_\mathbf{k}\rangle \langle n_\mathbf{k}| \partial_{k_j}n_\mathbf{k} \rangle
	-\langle \partial_{k_j} n_\mathbf{k} | n_\mathbf{k}\rangle \langle n_\mathbf{k}| \partial_{k_i}n_\mathbf{k} \rangle
	\right]. \label{fsdef}
\end{align}
Note that $\Omega_{ij}^n$ is anti-symmetric in the indices $i$ and $j$, while $g_{ij}^n$ is symmetric. 
Both the Berry curvature and the Fubini-Study metric are geometrical properties of the energy band and are invariant under the gauge transformation $|n_\mathbf{k}\rangle \to e^{i\theta(\mathbf{k})}|n_\mathbf{k}\rangle$. Both quantities can be written in a uniform manner in terms of the quantum geometric tensor defined by~\cite{Kolodrubetz:2016}
\begin{align}
	\chi_{ij}^n(\mathbf{k}) \equiv \langle \partial_{k_i} n_\mathbf{k} |(1 - | n_\mathbf{k}\rangle \langle n_\mathbf{k}|)| \partial_{k_j}n_\mathbf{k} \rangle.
\end{align}
Its real part gives the Fubini-Study metric $g_{ij}^n = \mathrm{Re}\,\chi_{ij}^n$, and the imaginary part gives the Berry curvature $\Omega_{ij}^n = -2\mathrm{Im}\,\chi_{ij}^n$.
The goal of this paper is to find a simple relation between the quantum geometric tensor and observable quantities in the steady-state of driven-dissipative systems and give a concrete protocol to experimentally determine the quantum geometric tensor.

\textit{Steady-state Hall response.}
The Hamiltonian of a generic non interacting tight-binding lattice model takes a quadratic form $\hat{H} = \sum_{\mathbf{r},\mathbf{r}^\prime} \hat{b}_\mathbf{r}^\dagger H_{\mathbf{r},\mathbf{r}^\prime} \hat{b}_{\mathbf{r}^\prime}$, where $\hat{b}_\mathbf{r}^\dagger$ and $\hat{b}_\mathbf{r}$ are the creation and annihilation operators of a boson at site $\mathbf{r}$. The corresponding Heisenberg equations of motion are a set of classical linear equations, after taking the expectation values of operators: $i\partial_t \beta_\mathbf{r} (t)= \sum_{\mathbf{r}^\prime} H_{\mathbf{r},\mathbf{r}^\prime} \beta_{\mathbf{r}^\prime}(t)$, where $\beta_{\mathbf{r}}(t) \equiv \langle \hat{b}_\mathbf{r} (t)\rangle$. We assume that the system has a uniform loss (dissipation) of rate $\gamma$ from each site, which can be incorporated into the Heisenberg equations of motion by adding imaginary diagonal terms $-i\gamma \beta_\mathbf{r}(t)$ to the right-hand side.
Then, the Heisenberg equation of motion describing the driven-dissipative lattice system is~\cite{Ozawa:2014, Bardyn:2014}
\begin{align}
	i\partial_t |\beta (t)\rangle
	=
	\left( H - i\gamma \mathbb{I}_{N} \right) |\beta(t) \rangle + |f(t)\rangle,
\end{align}
where $H$ is the matrix whose $\mathbf{r}$-$\mathbf{r}^\prime$ component is $H_{\mathbf{r},\mathbf{r}^\prime}$, $|\beta (t)\rangle$ is a vector whose components are $\beta_\mathbf{r}(t)$, $|f(t)\rangle$ is a vector describing the driving term, and $\mathbb{I}_{N}$ is an $N$-by-$N$ identity matrix with $N$ being the number of lattice sites in the system.
This equation is fully classical; it not only describes non interacting quantum mechanical systems but also classical light in coupled resonators~\cite{Hafezi:2011,Hafezi:2013,Anderson:2016} and coupled classical mechanical pendula~\cite{Salerno:2016}.

Assuming that the driving is monochromatic with the frequency $\omega_0$, $|f(t)\rangle = |f\rangle e^{-i\omega_0 t}$, there exists a steady state of the system which oscillates with the same frequency $|\beta (t)\rangle = |\beta \rangle e^{-i\omega_0 t}$. The steady state is experimentally obtained by driving the system and waiting for a sufficiently long time compared to $1/\gamma$.
The steady-state wavefunction is obtained by solving the following time-independent linear equation:
\begin{align}
	\left\{(\omega_0 + i\gamma)\mathbb{I}_{N} - H \right\} |\beta\rangle = |f\rangle. \label{lineareq}
\end{align}
We now look at an analog of the quantum Hall effect: to apply a small force $E_j$ in one direction, $r_j$, and to see a response in the center of mass in another direction, $r_i$. The force is applied by adding a term $-\hat{r}_j E_j$ in the Hamiltonian, where $\hat {r}_j$ is the position operator in the $r_j$ direction.

To understand how the center-of-mass position can depend on the geometry in momentum space, we Fourier transform the state $|\beta\rangle$ and the drive $|f\rangle$, which we write $|\beta(\mathbf{k})\rangle$ and $|f(\mathbf{k})\rangle$, respectively. The Fourier transformation of each sublattice component is taken separately; practically, $|\beta(\mathbf{k})\rangle$ and $|f(\mathbf{k})\rangle$ are vectors with $q$ components.
In order to obtain a simple relation between the center of mass and the momentum-space geometry, we consider performing $q$ different measurements. Looking at each unit cell, the driving fields corresponding to these $q$ measurements should form an orthonormal set of vectors. For example, we drive only one sublattice for each measurement and going through all the $q$ different sublattices with $q$ measurements. (We give the full details of the driving and detection scheme in the Supplemental Material~\cite{SupMat}.) More mathematically, we write $q$ different driving fields as $|f^l(\mathbf{k})\rangle$, with $l = 1,2,\cdots, q$, where each $|f^l(\mathbf{k})\rangle$ is a $q$-component vector for a given $\mathbf{k}$. We assume that $|f^l(\mathbf{k})\rangle$ can be separated into $\mathbf{k}$-dependent real-valued profile function $f(\mathbf{k})$ and the $\mathbf{k}$-independent sublattice structure as $|f^l(\mathbf{k})\rangle = f(\mathbf{k}) |l\rangle$. The orthonormality within a unit cell means that the $q$-component vectors $|l\rangle$ satisfy the completeness condition $\sum_{l=1}^q |l\rangle \langle l | = \mathbb{I}_q$.

We denote the steady-state field amplitudes in real and momentum spaces when the driving field is $|f^l (\mathbf{k})\rangle$ by $|\beta^l\rangle$ and $|\beta^l (\mathbf{k})\rangle$, respectively. 
We then consider the following combined center-of-mass displacement:
\begin{align}
	\llangle \hat{r}_i \rrangle
	\equiv
	\frac{\sum_l \langle \beta^l | \hat{r}_i | \beta^l\rangle}{\sum_l \langle \beta^l | \beta^l\rangle}
	=
	\frac{\sum_l \sum_\mathbf{k}\langle \beta^l(\mathbf{k})|i\partial_{k_i}|\beta^l(\mathbf{k})\rangle}{\sum_l \sum_\mathbf{k} \langle \beta^l(\mathbf{k})|\beta^l(\mathbf{k})\rangle}
	\equiv
	\frac{\llangle \hat{r}_i \rrangle_\mathrm{num}}{\llangle \hat{r}_i \rrangle_\mathrm{den}},
	\label{xidef}
\end{align}
which is an experimentally measurable quantity.
We defined the numerator as $\llangle \hat{r}_i \rrangle_\mathrm{num}$ and the denominator as $\llangle \hat{r}_i \rrangle_\mathrm{den}$.
Previously, we analyzed the structure of $\llangle r_i\rrangle$ for specific models when $f(\mathbf{k})$ is constant and derived how the Chern number or the energy-resolved Berry curvature can be extracted from the steady state~\cite{Ozawa:2014,Ozawa:2016,Salerno:2016}.
Now we consider general lattice models with a general driving $|f(\mathbf{k})\rangle$ and reveal the dependence of $\llangle \hat{r}_i\rrangle$ on the full quantum geometric tensor. In the presence of a force, 
the steady state $|\beta^l (\mathbf{k})\rangle$, up to the first order in $E_j$, is 
\begin{align}
	&|\beta^l(\mathbf{k})\rangle
	=
	\frac{1}{(\omega_0 + i\gamma)\mathbb{I}_{q} - H_\mathbf{k}}
	|f^l(\mathbf{k})\rangle
	\notag \\
	&-
	\frac{i E_j}{(\omega_0 + i\gamma)\mathbb{I}_{q} - H_\mathbf{k}}
	\frac{\partial}{\partial k_j}
	\frac{1}{(\omega_0 + i\gamma)\mathbb{I}_{q} - H_\mathbf{k}}
	|f^l(\mathbf{k})\rangle, \label{betal}
\end{align}
where $H_\mathbf{k}$ is the momentum-space Hamiltonian in the absence of the force.

Using (\ref{betal}), the combined center of mass $\llangle \hat{r}_i\rrangle$ up to the lowest order in the applied force can be calculated. We give the full expression of $\llangle \hat{r}_i\rrangle$ for a general lattice in the Supplemental Material~\cite{SupMat}, and in the main text we focus on the situation where the band of interest is flat enough and sufficiently separated from other bands. We set the frequency of the driving $\omega_0$ to be close to the $n$th band whose energy is $\mathcal{E}_n$ and define the detuning $\Delta \equiv \omega_0 - \mathcal{E}_n$. In general, even if the band is not completely flat, our results hold if the detuning $\Delta$ and the loss $\gamma$ is sufficiently larger than the bandwidth, but sufficiently smaller than the bandgap to the neighboring bands. Under these conditions, the denominator and the numerator of (\ref{xidef}) are
\begin{align}
	&\llangle \hat{r}_i \rrangle_{\mathrm{den}}
	=
	\frac{1}{\Delta^2 + \gamma^2}\sum_\mathbf{k} f(\mathbf{k})^2
	+
	\mathcal{O}(E_j^2),
	\notag \\
	&\llangle \hat{r}_i \rrangle_\mathrm{num}
	=
	-\frac{E_j}{(\Delta^2 + \gamma^2)^2}
\sum_\mathbf{k}
	\left\{
	\gamma f(\mathbf{k})^2 \Omega^n_{ij} (\mathbf{k})
	\right.
	\notag \\
	&\left.
	+
	2\Delta
	\left(
	f(\mathbf{k})^2 g^n_{ij} (\mathbf{k})
	+ \partial_{k_i}f(\mathbf{k}) \partial_{k_j}f(\mathbf{k})
	\right)
	\right\}
	+\mathcal{O}(E_j^2).
\end{align}
We can see that the combined center of mass $\llangle \hat{r}_i \rrangle = \llangle \hat{r}_i \rrangle_\mathrm{num}/\llangle \hat{r}_i \rrangle_\mathrm{den}$ depends on both the Berry curvature and the Fubini-Study metric.

If we choose the driving field profile $f(\mathbf{k})$ to be independent of $\mathbf{k}$, the combined center of mass $\llangle \hat{r}_i \rrangle$ depends on the average of the Berry curvature and the Fubini-Study metric, and for the Harper-Hofstadter model, one recovers the results obtained in Refs.~\cite{Ozawa:2014,Ozawa:2016,Salerno:2016}.

On the other hand, using a driving field which has the Gaussian form, one can access the quantum geometric tensor with resolution in momentum space as we now show.
We assume that the driving field profile takes the following Gaussian form
$
	f(\mathbf{k})
	=
	f \exp \left( -(\mathbf{k} - \mathbf{k}_0)^2 / 2\sigma_k^2\right)$,
where $\mathbf{k}_0$ is the center of the Gaussian in momentum space and $\sigma_k$ is the spread of the Gaussian.
Assuming that $\sigma_k$ is small enough so that the change in $f(\mathbf{k})$ around $\mathbf{k}_0$ is much faster than that in $\Omega_{ij}^n(\mathbf{k})$ and $g_{ij}^n(\mathbf{k})$, we obtain the following expression:
\begin{align}
	\llangle \hat{r}_i\rrangle
	\approx
	-\frac{E_j}{\Delta^2 + \gamma^2}
	\left\{
	\gamma \Omega_{ij}^n (\mathbf{k}_0)
	+ 2\Delta \left( g_{ij}^n (\mathbf{k}_0) + \frac{\delta_{ij}}{2 \sigma_k^2}\right)
	\right\}, \label{mainresult}
\end{align}
where $\delta_{ij}$ is the Kronecker delta.
This expression is the central result of the paper, which relates the steady-state response $\llangle \hat{r}_i \rrangle$ and the full quantum geometric tensor.
This expression, in turn, allows one to estimate the momentum-resolved full quantum geometric tensor from the driven-dissipative steady state.
We did not specify the dimensionality of the lattice; this expression is valid in any dimension including four spatial dimensions~\cite{Ozawa:2016, Price:2016}.
Restricting to two dimensions, one can extract the quantum geometry in the following way.
To estimate the only nonzero component of the Berry curvature, $\Omega_{xy}^n = -\Omega_{yx}^n$, one can set the detuning $\Delta$ to be zero, and $\Omega_{xy}^n (\mathbf{k}_0) \approx -\llangle \hat{x}\rrangle \gamma/E_y$.
To estimate $xy$ component of the Fubini-Study metric, $g_{xy}^n = g_{yx}^n$, one can perform two sets of measurements; add a force in $y$ direction and measure $\llangle \hat{x} \rrangle$, and add a force in $x$ direction and measure $\llangle \hat{y} \rrangle$. This gives
$g_{xy}^n (\mathbf{k}_0) \approx -\frac{\Delta^2 + \gamma^2}{4\Delta}\left[\frac{\llangle \hat{x} \rrangle}{E_y} + \frac{\llangle \hat{y} \rrangle}{E_x} \right]$.
To estimate the diagonal component of the Fubini-Study metric, e.g., $g_{xx}^n$, one can add a force in $x$ direction and measure $\llangle \hat{x} \rrangle$ which gives $g_{xx}^n (\mathbf{k}_0) \approx -\llangle \hat{x} \rrangle \frac{\Delta^2 + \gamma^2}{2\Delta E_x} - \frac{1}{2\sigma_k^2}$.
We can thus determine all the components of the quantum geometric tensor.
Next, we numerically demonstrate this scheme using the two-dimensional Lieb lattice as an example.

\begin{figure}[htbp]
\begin{center}
\subfigure[Two-dimensional Lieb lattice]{
\includegraphics[width= 0.23 \textwidth]{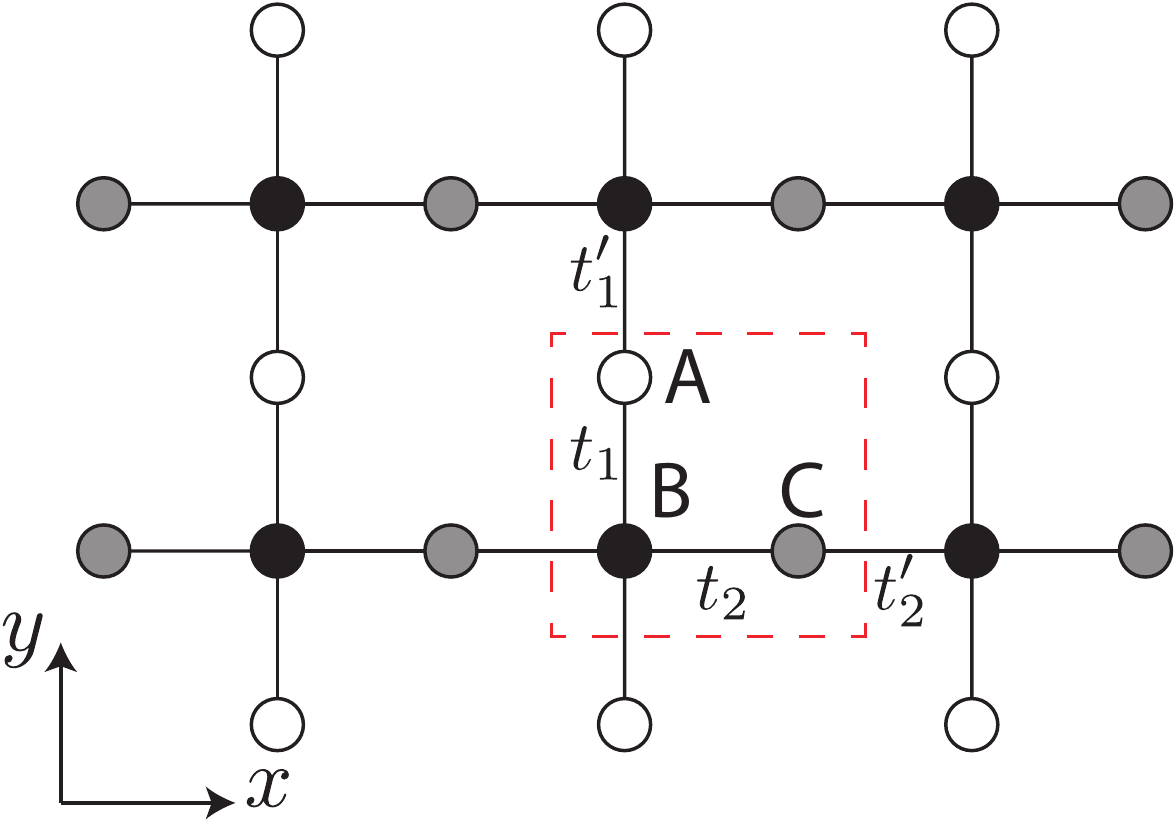}}
\subfigure[Energy dispersion]{
\includegraphics[width= 0.23 \textwidth]{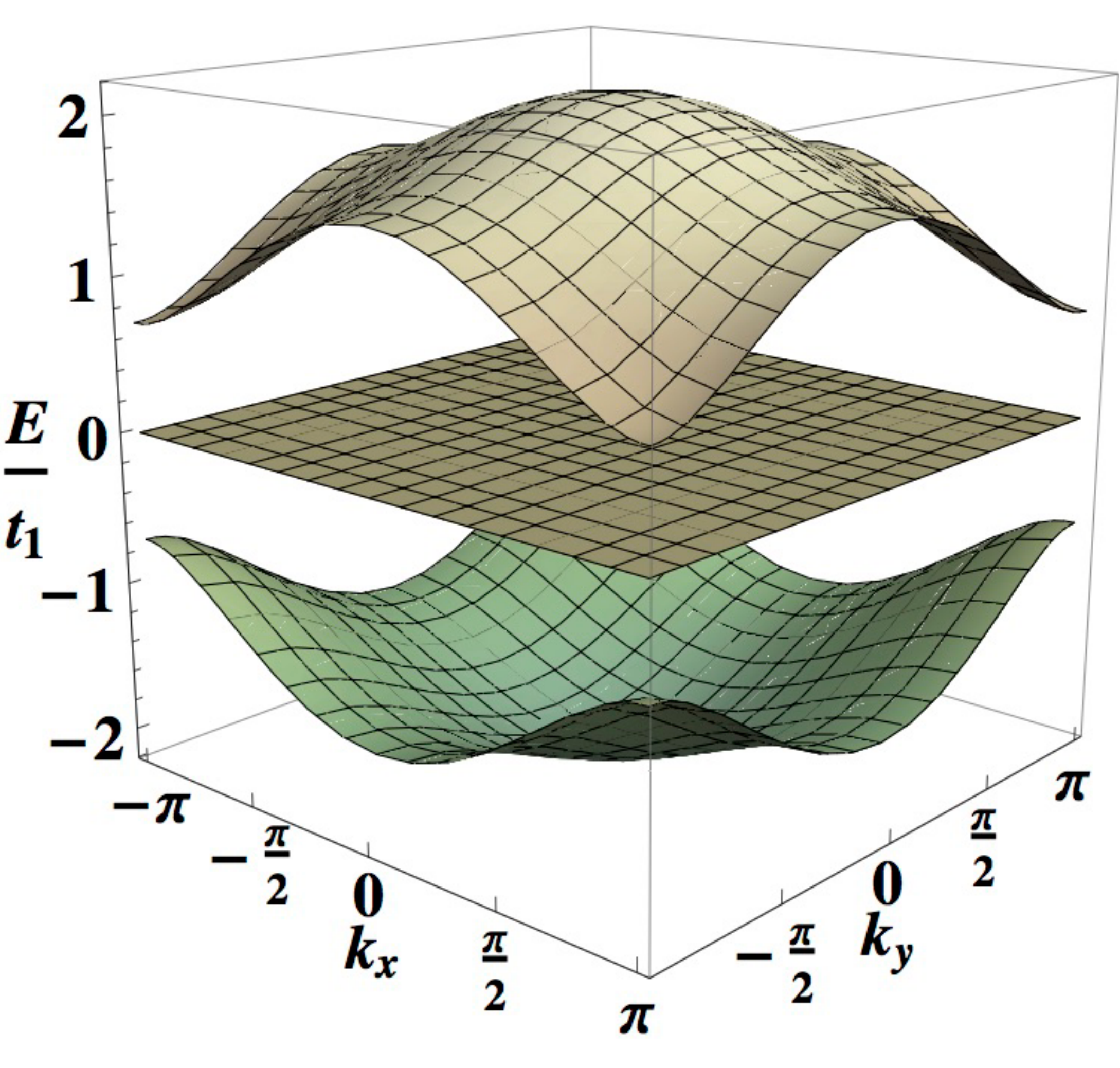}}
\caption{(a) A schematic structure of the two-dimensional Lieb lattice. Three sublattices, A, B, and C, are indicated respectively by white, black, and gray circles. Three lattice sites, such as the ones indicated within the dashed square, form a unit cell, and the respective hopping amplitudes are indicated. (b) The dispersion of the two-dimensional Lieb lattice as a function of momenta $k_x$ and $k_y$ with parameters $t_1 = t_2 = 2 t_1^\prime = 2 t_2^\prime$.}
\label{2dlieb}
\end{center}
\end{figure}

\textit{Two-dimensinal (2D) Lieb lattice.}
The two-dimensional Lieb lattice is a tight-binding model with three sites per unit cell, as described in Fig.~\ref{2dlieb}~\cite{Guzman:2014, Poli:2017}. A unit cell contains three lattice sites belonging to three sublattices, A, B, and C. Four different types of hopping amplitudes, $t_1$, $t_1^\prime$, $t_2$, and $t_2^\prime$ exist, which are indicated in the figure. 
The tight-binding Hamiltonian of the 2D Lieb lattice is
\begin{align}
	H = -\sum_{x,y}
	&\left( t_1 a_{x,y}^\dagger b_{x,y} + t_2 c_{x,y}^\dagger b_{x,y}
	\right.	
	\notag \\
	&\left.+ t_2^\prime b_{x+1,y}^\dagger c_{x,y} + t_1^\prime b_{x,y+1}^\dagger a_{x,y} + \mathrm{H.c.}\right),
\end{align}
where $a_{x,y}$ is the annihilation operator of a particle in the A sublattice in the unit cell located at position $(x,y)$. The annihilation operators $b_{x,y}$ and $c_{x,y}$ are defined similarly. We take the spacing between the adjacent unit cells to be unity.
The model has three bands, and the middle band is flat with zero energy.
A typical dispersion of the model is plotted in Fig.~\ref{2dlieb}(b).

\begin{figure*}[htbp]
\begin{center}
\subfigure[Estimated Berry curvature $\Omega_{xy}(k_x,k_y)$]{
\includegraphics[width= 0.23 \textwidth]{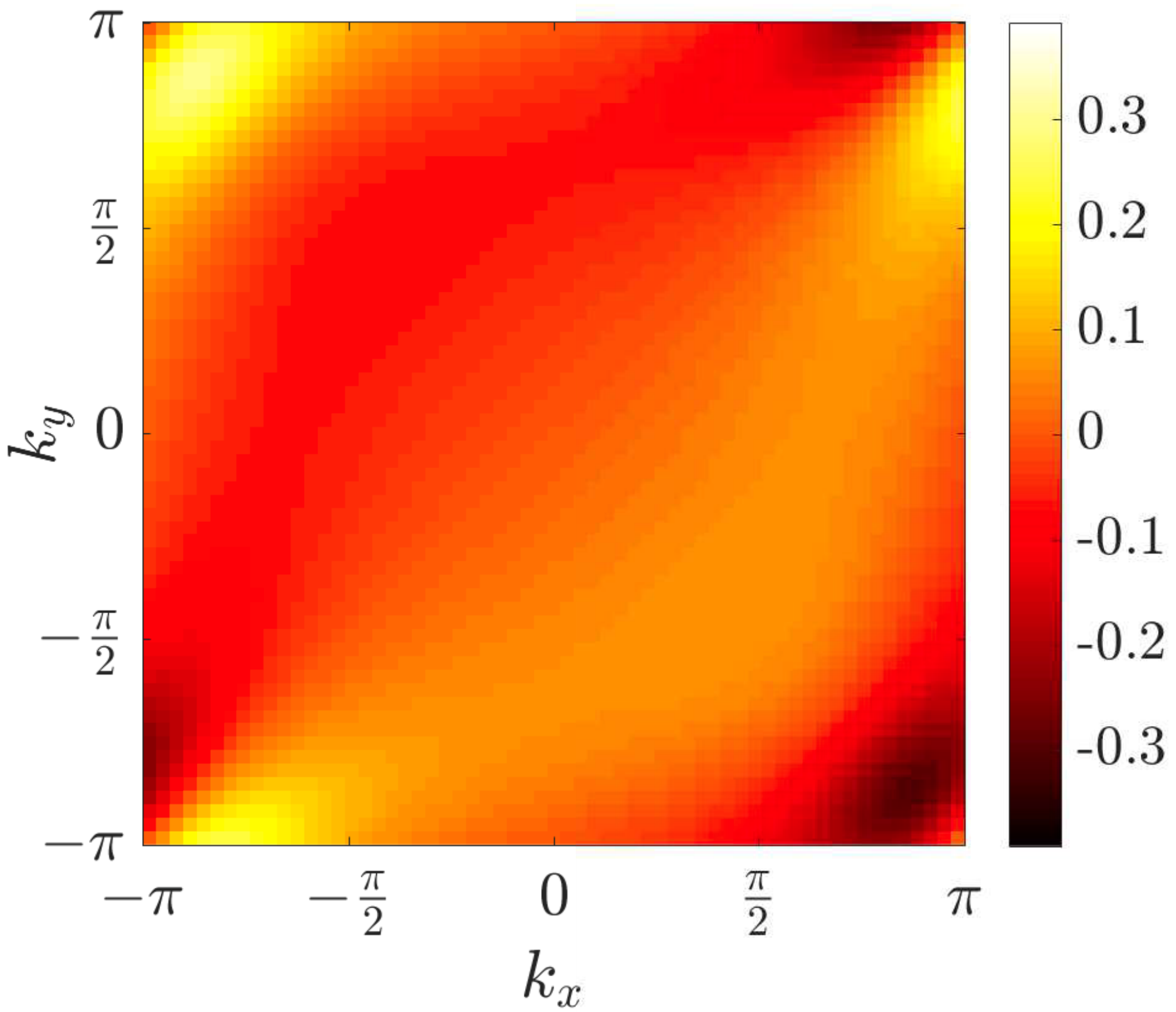}}
\subfigure[Estimated Fubini-Study $g_{xy}(k_x,k_y)$]{
\includegraphics[width= 0.23 \textwidth]{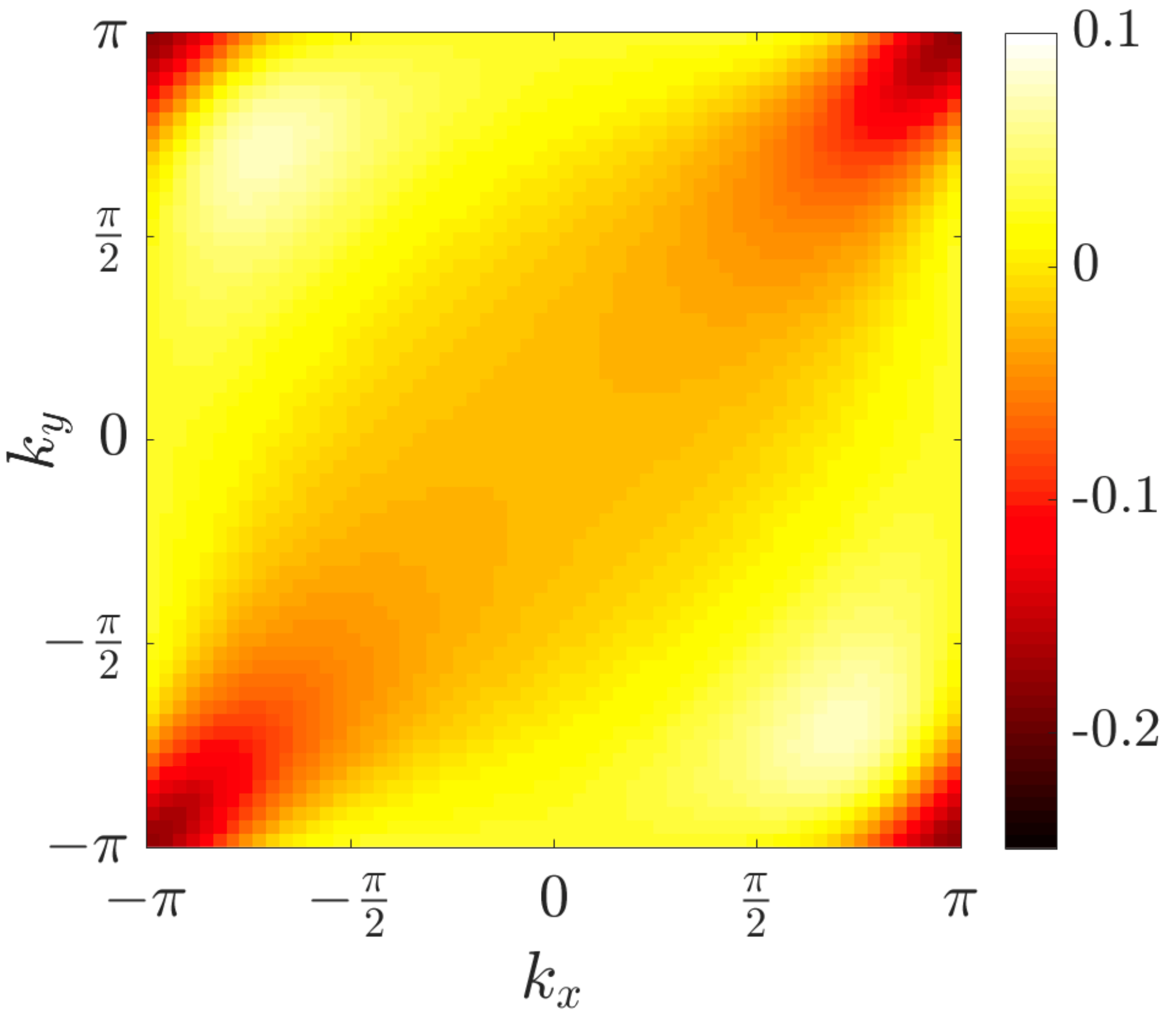}}
\subfigure[Estimated Fubini-Study $g_{xx}(k_x,k_y)$]{
\includegraphics[width= 0.23 \textwidth]{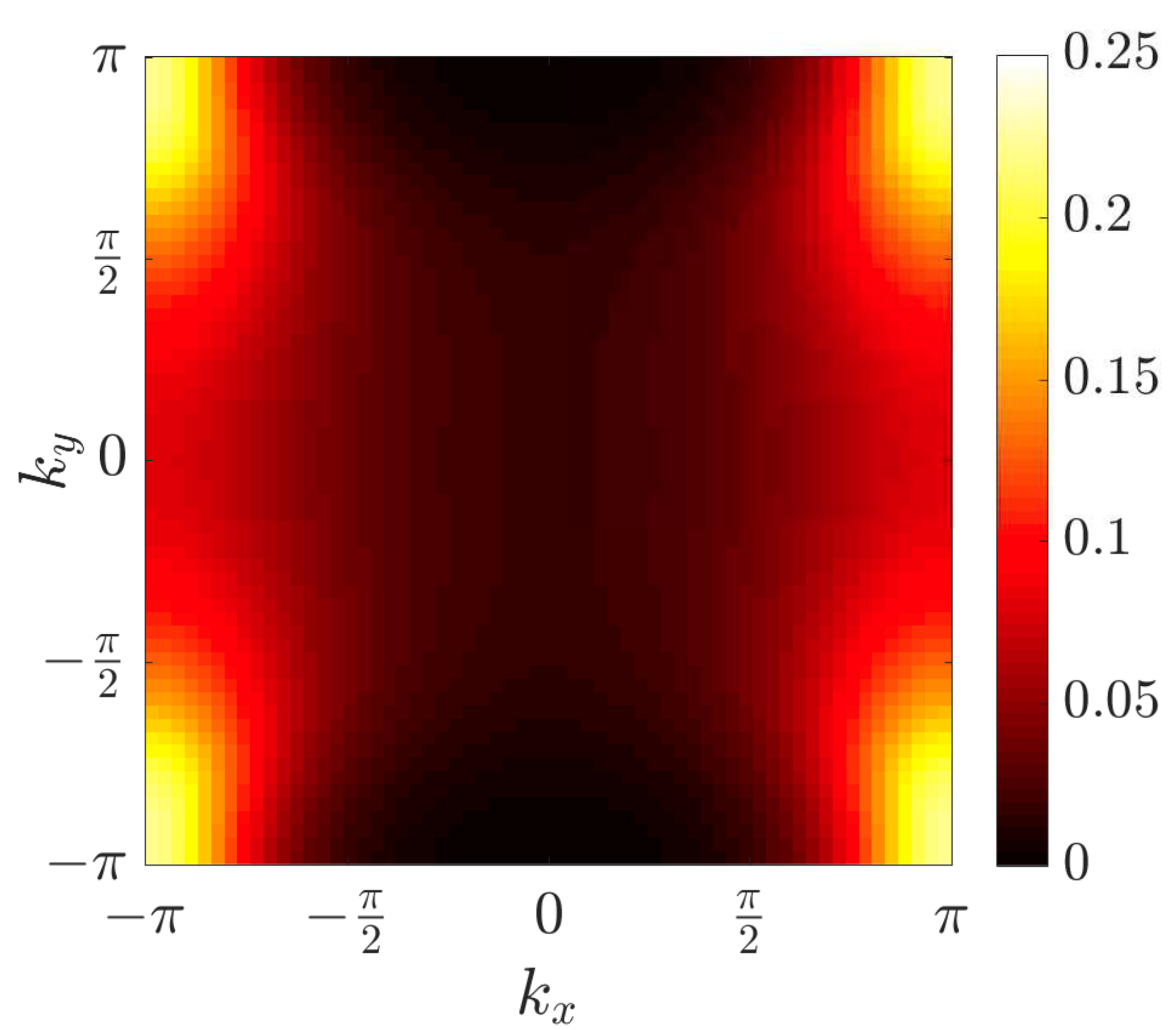}}
\subfigure[Estimated Fubini-Study $g_{yy}(k_x,k_y)$]{
\includegraphics[width= 0.23 \textwidth]{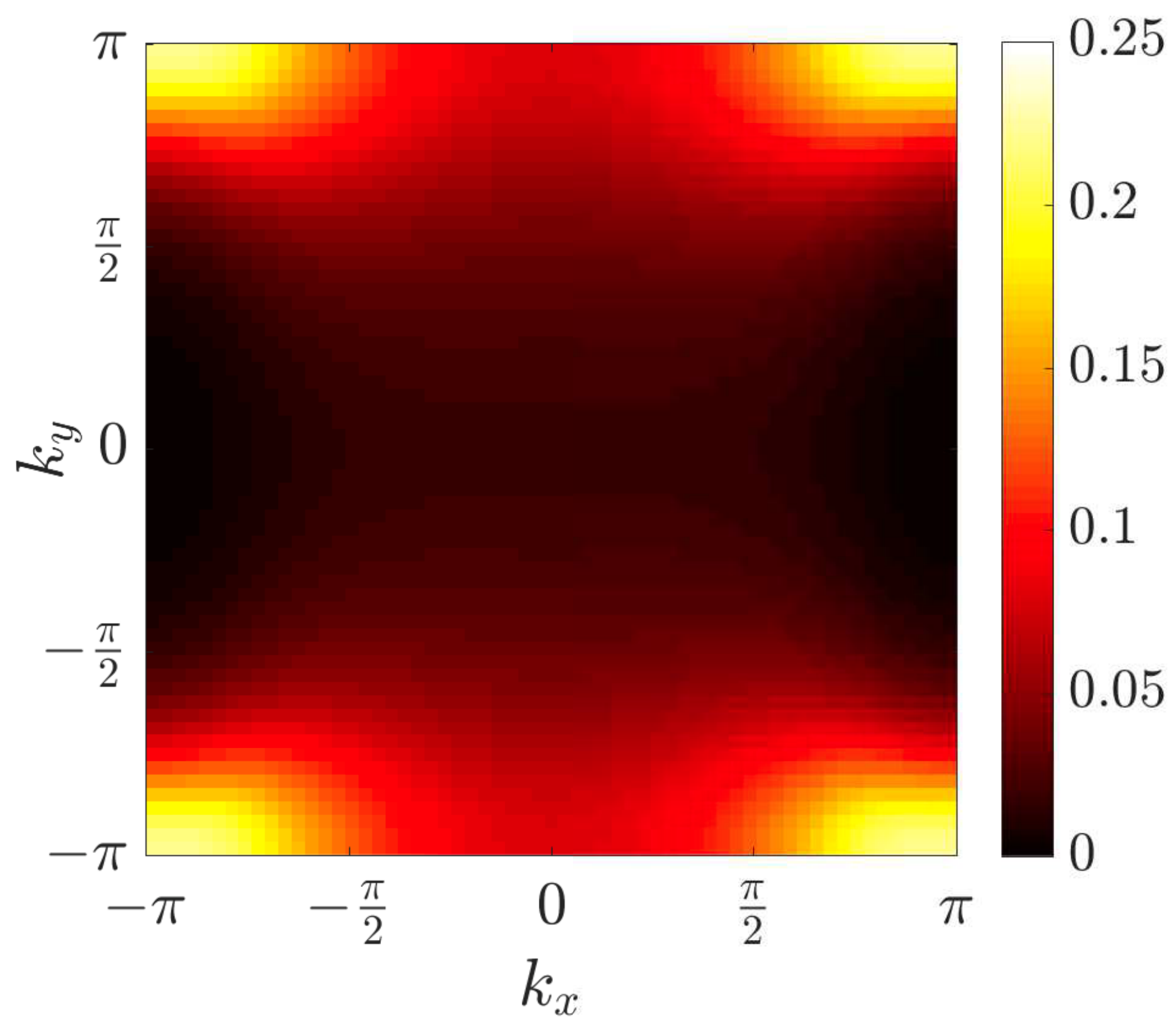}}
\subfigure[Ideal Berry curvature $\Omega_{xy}(k_x,k_y)$]{
\includegraphics[width= 0.23 \textwidth]{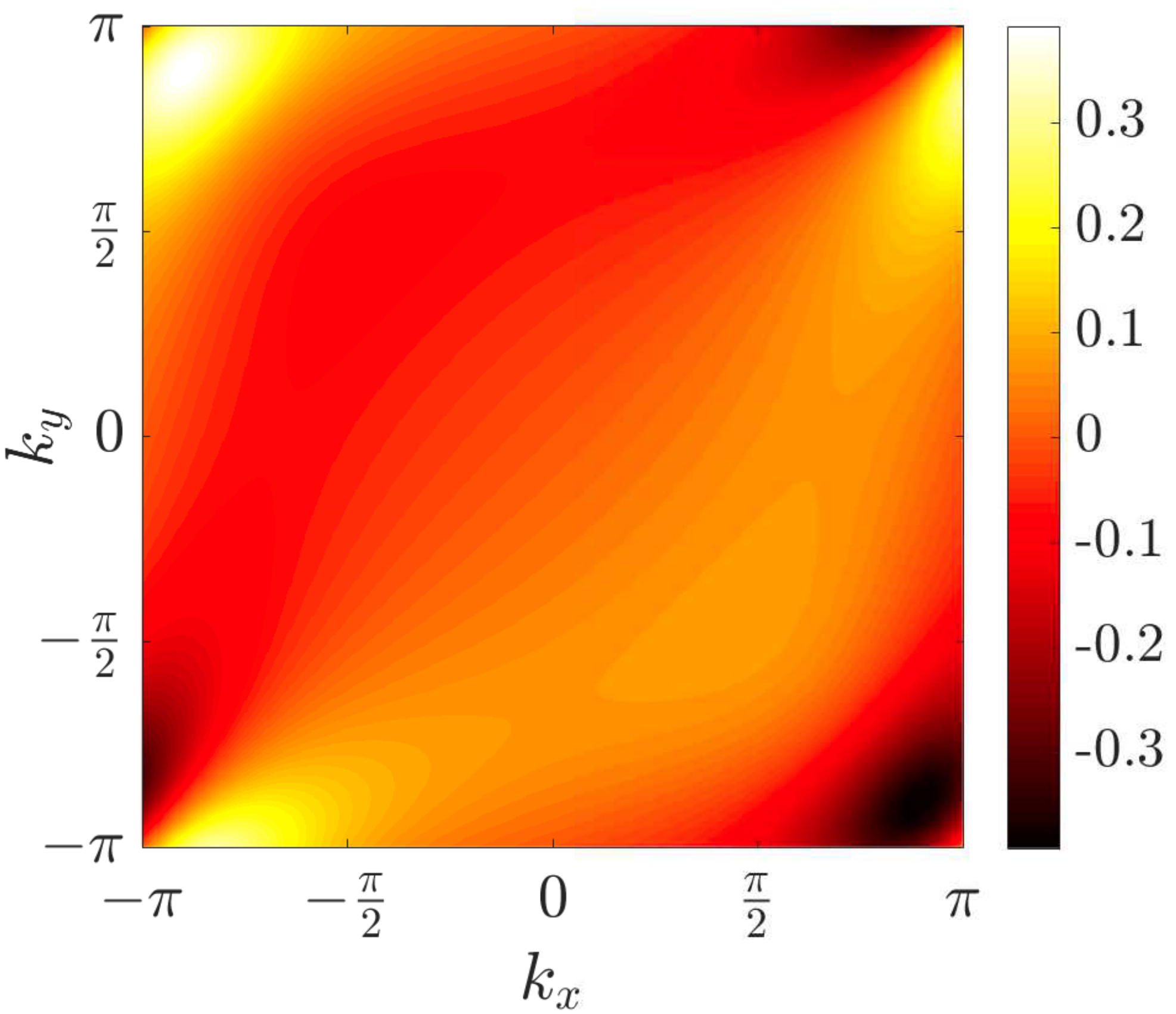}}
\subfigure[Ideal Fubini-Study $g_{xy}(k_x,k_y)$]{
\includegraphics[width= 0.23 \textwidth]{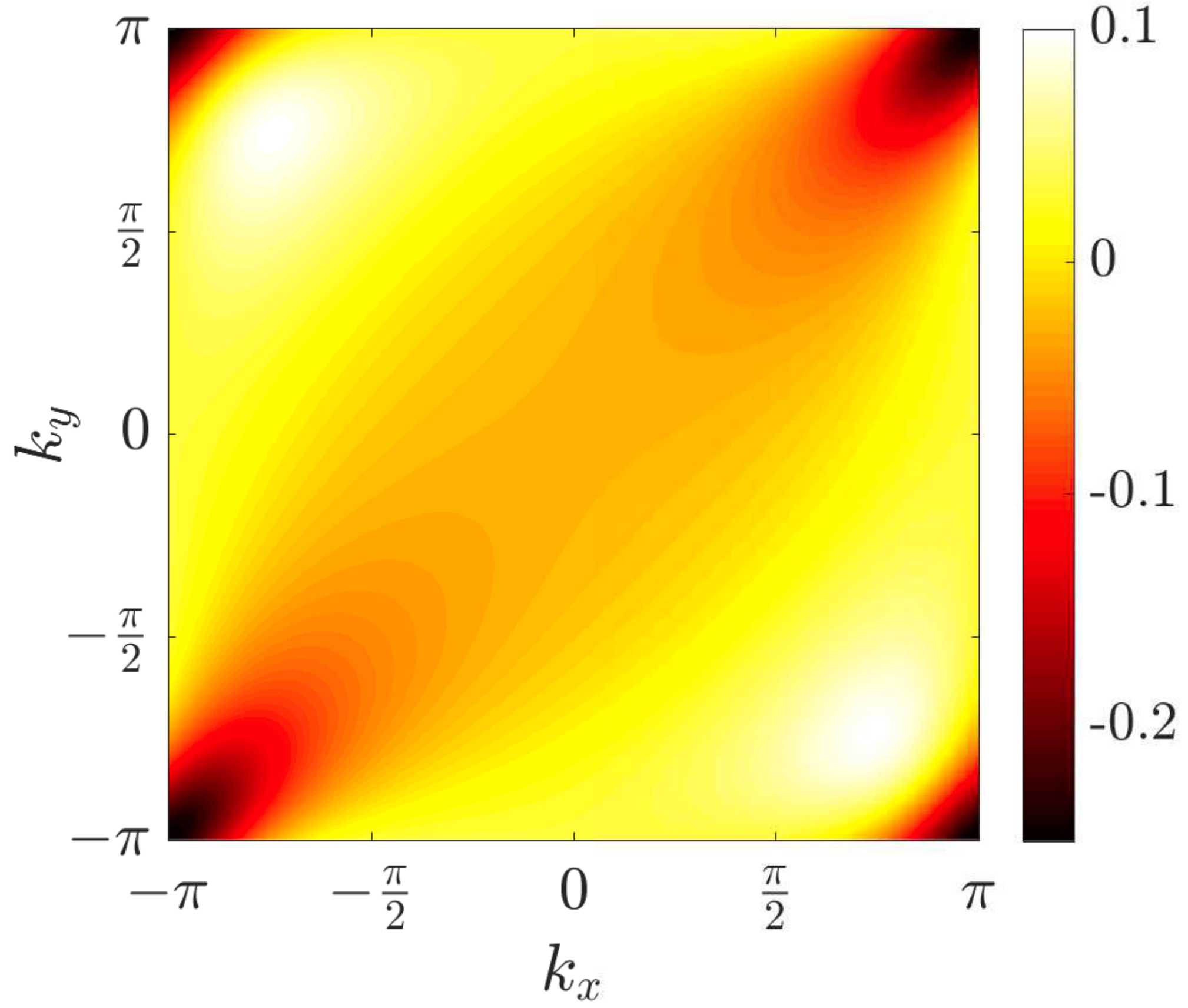}}
\subfigure[Ideal Fubini-Study $g_{xx}(k_x,k_y)$]{
\includegraphics[width= 0.23 \textwidth]{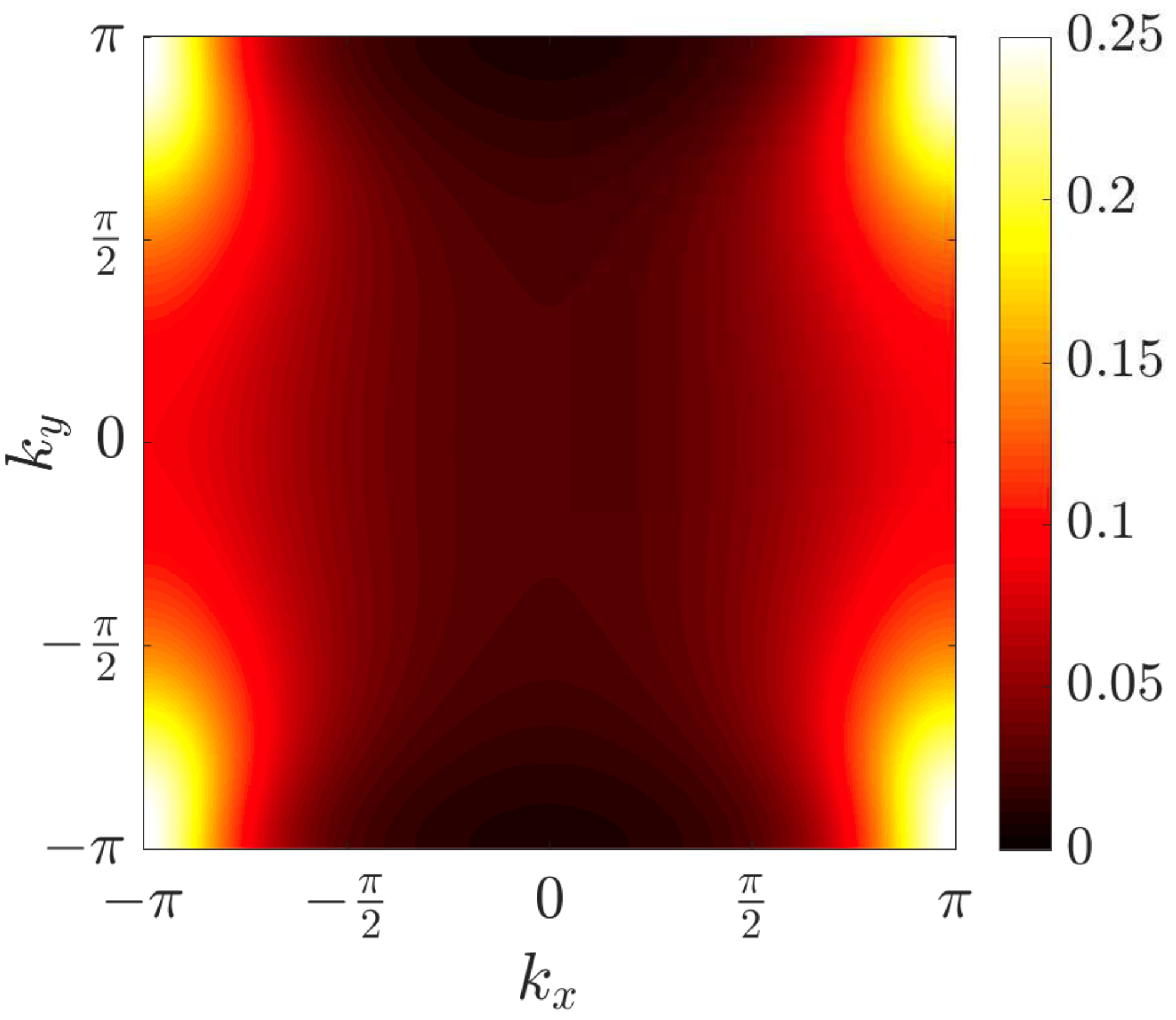}}
\subfigure[Ideal Fubini-Study $g_{yy}(k_x,k_y)$]{
\includegraphics[width= 0.23 \textwidth]{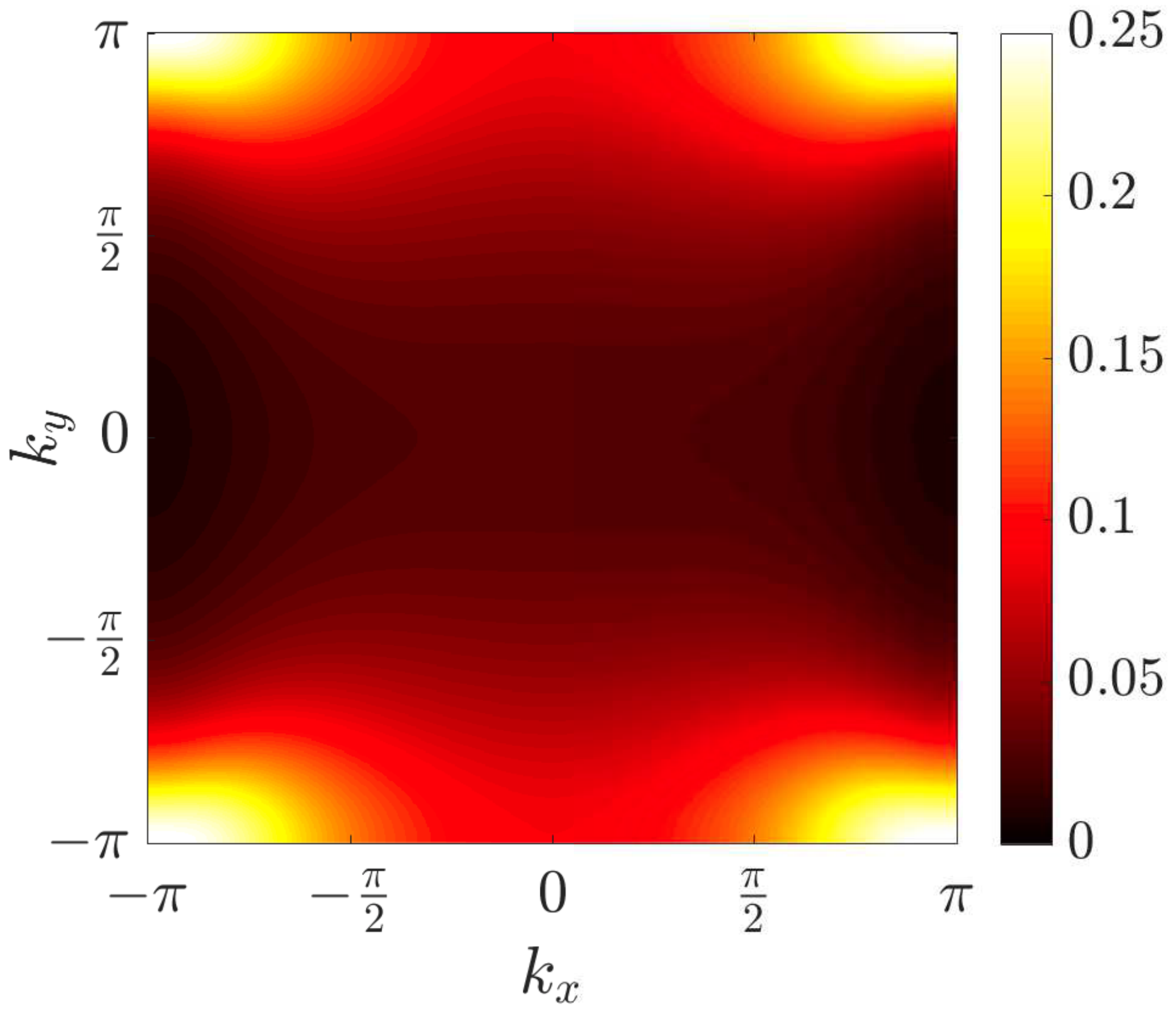}}
\caption{Complete quantum geometric tensor of a 2D Lieb lattice. [(a)-(d)] Estimated values from the numerical simulation of the driven-dissipative steady-state using (\ref{mainresult}); the left-hand side of (\ref{mainresult}) is estimated from the full numerical simulation of (\ref{lineareq}) and then (\ref{xidef}), and we use (\ref{mainresult}) to extract the components of the quantum geometric tensor. [(e)-(h)] Ideal values calculated from the definition of the Berry curvature (\ref{berrydef}) and the Fubini-Study metric (\ref{fsdef}). All the figures are in units of the lattice spacing squared.}
\label{geometry}
\end{center}
\end{figure*}

We now apply the method developed in the previous section to numerically estimate the quantum geometry of the flat band. We consider a lattice of $40 \times 40$ unit cells, and we take the lattice parameters to be $t_1 = t_2 = 2 t_1^\prime = 2 t_2^\prime$, corresponding to the band gap of $\sim 0.7t_1$ to the other bands. For all the calculations, we assume that the driving field has the width $\sigma_k = 1/5$.
Upon numerically estimating $\llangle \hat{r}_i\rrangle$ up to the first order in $E_j$, we perform differential measurements with opposite forces $E_j$ and $-E_j$ in order to remove contributions from second (and any even) order terms in $E_j$.

To map out the $xy$ component of the Berry curvature $\Omega_{xy}$ and the Fubini-Study metric $g_{xy}$ of the middle band, we use the detunings $\Delta = 0$ and $\Delta = 0.1 t_1$, respectively, and the loss $\gamma = 0.1t_1$ and the force $E_j = 0.01 t_1$.
Numerically estimated $\Omega_{xy}$ and $g_{xy}$ are plotted in Figs.~\ref{geometry}(a) and~\ref{geometry}(b), together with their ideal values calculated according to their definitions in Figs.~\ref{geometry}(e) and~\ref{geometry}(f). The agreement we find is excellent. The parameters we use here are within the current state-of-the art experiments of photonic lattices~\cite{Jacqmin:2014}.

Estimating the diagonal ($xx$ and $yy$) components of the Fubini-Study metric requires more stringent conditions. This is because, in Eq.~(\ref{mainresult}), the term $1/2\sigma_k^2$ dominates over $g_{xx}$ or $g_{yy}$ and thus the Fubini-Study metric contribution in $\llangle \hat{r}_i \rrangle$ becomes a small correction over $1/2\sigma_k^2$.
In order to correctly capture this small correction, we need to use smaller values of $\Delta$ and $\gamma$ (otherwise the effects of other bands enter) and also a small force (otherwise third or higher order terms in $E_j$ enter).
In  Figs.~\ref{geometry}(c) and~\ref{geometry}(d), we plot the estimated values of $g_{xx}$ and $g_{yy}$ using $\Delta = 0.01t_1$, $\gamma = 0.01t_1$, and $E_j = 0.0001t_1$. The ideal values of the Fubini-Study metric are plotted in Figs.~\ref{geometry}(g) and~\ref{geometry}(h), which show good agreement with the estimated values. For diagonal terms of the Fubini-Study metric, the necessity in using small parameters is a major challenge in experimentally implementing the scheme. In general, having a larger band gap allows one to use larger values of the parameters.

\textit{Conclusion.}
We have shown that a steady-state Hall response of bosons in driven-dissipative lattices has a simple dependence on the full quantum geometric tensor. The relation we find is valid for lattices with any spatial dimensions. We have demonstrated how one can estimate the momentum-resolved quantum geometric tensor using the steady-state Hall response for the flat band of the 2D Lieb lattice. Our method is of direct experimental relevance in various synthetic photonic systems, such as the silicon micro-ring resonators~\cite{Hafezi:2011, Hafezi:2013}, exciton-polariton microcavity arrays~\cite{Jacqmin:2014, Baboux:2016, Whittaker:2017}, and microwave cavity arrays~\cite{Anderson:2016}.
The generalization of the method to extract the full quantum geometric tensor for non flat bands are left for future works.
We have focused on the case where the emission from each cavity can be isolated and measured independently. It would be interesting to extend the method to a more general situation where the emission from nearby cavities overlap.

Our method opens a unique way for bosonic topological systems, such as photonic~\cite{Lu:2014, Lu:2016} or mechanical systems~\cite{Susstrunk:2015, Huber:2016}, to explore geometrical and topological band structures with probes not available in electronic topological insulators. We hope that our work stimulates the further study, both theoretically and experimentally, on the understanding of the quantum geometric tensor.

\begin{acknowledgments}
The author is grateful to Iacopo Carusotto and Hannah Price for frequent discussions since the early stage of the work, and Grazia Salerno for careful reading of the manuscript. The author also thanks Olivier Bleu, Ajit Srivastava, Victor Albert, and Fr\'ed\'eric Pi\'echon for stimulating discussions and exchanges.
This work was funded by the EU-FET Proactive grant AQuS, Project No. 640800.
\end{acknowledgments}

\clearpage
\renewcommand{\theequation}{S\arabic{equation}}. \setcounter{equation}{0}

\makeatletter
\renewcommand\@biblabel[1]{[S#1]}
\makeatother

\begin{widetext}

\begin{center}
\begin{large}
{\bf Supplemental Material for ``Steady-state Hall response and quantum geometry of driven-dissipative lattices"}
\end{large}
\end{center}

\section{I. Details of driving and detection schemes}
We describe here details of the driving and detection schemes we use.
The state of the system and the driving field are described by vectors which are denoted by $|\beta\rangle$ and $|f\rangle$, respectively. To understand the structures of $|\beta\rangle$ and $|f\rangle$ in a more intuitive manner, let's take a concrete example of the two-dimensional Lieb lattice introduced in the main text. A unit cell of the two-dimensional Lieb lattice has three sublattice sites, called A, B, and C sublattices.
Let $|A,\mathbf{r}\rangle$, $|B,\mathbf{r}\rangle$, and $|C,\mathbf{r}\rangle$ denote states with a particle at A, B, and C sublattices, respectively, in a unit cell positioned at $\mathbf{r}$.
Then, using these states as basis states, the state vector $|\beta\rangle$ can be expanded as
\begin{align}
	|\beta\rangle
	=
	\sum_{\mathbf{r}}
	\left( \beta_{A,\mathbf{r}} |A,\mathbf{r}\rangle + \beta_{B,\mathbf{r}} |B,\mathbf{r}\rangle + \beta_{C,\mathbf{r}} |C,\mathbf{r}\rangle\right),
\end{align}
where $\beta_{A,\mathbf{r}}$, $\beta_{B,\mathbf{r}}$, and $\beta_{C,\mathbf{r}}$ are, in general, complex numbers.
We define the Fourier transform $|\beta (\mathbf{k})\rangle$ of $|\beta\rangle$ separately for each sublattice.
For example, we perform the Fourier transform for A sublattice as
\begin{align}
	\beta_{A}(\mathbf{k})
	\equiv
	\frac{1}{\sqrt{N_c}}\sum_{\mathbf{r}} e^{-i\mathbf{k}\cdot \mathbf{r}}\beta_{A,\mathbf{r}},
\end{align}
where $N_c$ is the number of unit cells in the system. The relation $N = q N_c$ holds, where $N$ is the total number of lattice sites and $q$ is the number of lattice sites per unit cell. For the 2D Lieb lattice, $q = 3$.
The Fourier transforms $\beta_{B}(\mathbf{k})$ and $\beta_{C}(\mathbf{k})$ are defined similarly.
Writing the state with a particle only in sublattice A with momentum $\mathbf{k}$ as $|A,\mathbf{k}\rangle$, and defining $|B,\mathbf{k}\rangle$ and $|C,\mathbf{k}\rangle$ similarly, the Fourier transformed state vector $|\beta (\mathbf{k})\rangle$ is
\begin{align}
	|\beta(\mathbf{k})\rangle \equiv
	\beta_{A}(\mathbf{k})|A,\mathbf{k}\rangle + \beta_{B}(\mathbf{k})|B,\mathbf{k}\rangle + \beta_{C}(\mathbf{k})|C,\mathbf{k}\rangle.
\end{align}
Assuming that we always use $|A,\mathbf{k}\rangle$, $|B,\mathbf{k}\rangle$, and $|C,\mathbf{k}\rangle$ as basis states, we can conveniently write $|\beta(\mathbf{k})\rangle$ in a vector format as
\begin{align}
	|\beta(\mathbf{k})\rangle
	=
	\begin{pmatrix}
	\beta_{A}(\mathbf{k}) \\ \beta_{B}(\mathbf{k}) \\ \beta_{C}(\mathbf{k})
	\end{pmatrix}.
\end{align}
In general, $|\beta(\mathbf{k})\rangle$ is a $q$-component vector.

The driving field $|f\rangle$ and its Fourier transform $|f(\mathbf{k})\rangle$ are defined similarly. As described in the main text, we perform $q$ measurements with different driving fields $|f^l (\mathbf{k})\rangle$ and combine the results. We assume that $|f^l (\mathbf{k})\rangle$ can be decomposed into sublattice-independent overall profile $f(\mathbf{k})$ and the sublattice-dependent part as
\begin{align}
	|f^l (\mathbf{k})\rangle
	=
	\begin{pmatrix}
	f^l_{A}(\mathbf{k}) \\ f^l_{B}(\mathbf{k}) \\ f^l_{C}(\mathbf{k})
	\end{pmatrix}
	=
	f(\mathbf{k})
	|l\rangle.
\end{align}
Here, $|l\rangle$ is a $\mathbf{k}$-independent $q$-component vector. In order to simplify the expressions of the center of mass, we assume that $|l\rangle$ satisfies the completeness relation defined by
\begin{align}
	\sum_l |l\rangle \langle l| = \mathbb{I}_{q}.
\end{align}
This means that we need to perform measurements using $q$ sets of orthogonal vectors as driving fields.
We can achieve this by choosing, for example,
\begin{align}
	|1\rangle &= \begin{pmatrix}1 \\ 0 \\ 0\end{pmatrix}, &
	|2\rangle &= \begin{pmatrix}0 \\ 1 \\ 0\end{pmatrix}, &
	|3\rangle &= \begin{pmatrix}0 \\ 0 \\ 1\end{pmatrix},
\end{align}
when $q = 3$. This type of driving field requires driving only one sublattice at a time, which can be experimentally achieved by individual addressing of cavity sites, or by masking the sample covering sites except for one sublattice and performing experiments. Another choice, which is probably more experimentally straightforward, is to drive all sublattice sites but with different phases. Defining a nontrivial cubic root of 1 as $\omega \equiv e^{i2\pi/3}$, we can choose
\begin{align}
	|1\rangle &= \frac{1}{\sqrt{3}} \begin{pmatrix}1 \\ 1 \\ 1\end{pmatrix}, &
	|2\rangle &= \frac{1}{\sqrt{3}} \begin{pmatrix}1 \\ \omega \\ \omega^2\end{pmatrix}, &
	|3\rangle &= \frac{1}{\sqrt{3}} \begin{pmatrix}1 \\ \omega^2 \\ \omega\end{pmatrix},
\end{align}
which also satisfies the completeness relation.
Once we know the desired form of $|l\rangle$ and the profile function $f(\mathbf{k})$, the corresponding driving field in real space $|f^l\rangle$ can be obtained by Fourier transforming $f(\mathbf{k})$.
Namely, writing $|l\rangle = (l_A, l_B, l_C)^T$,
\begin{align}
	|f^l\rangle
	=
	\frac{1}{\sqrt{N_c}}\sum_{\mathbf{r}}
	\left(
	\sum_\mathbf{k} e^{i\mathbf{k}\cdot \mathbf{r}}f(\mathbf{k})
	\right)
	\left(	
	l_A |A,\mathbf{r}\rangle
	+
	l_B |B,\mathbf{r}\rangle
	+
	l_C |C,\mathbf{r}\rangle
	\right).
\end{align}
Defining the profile function $f(\mathbf{r})$ in real space as
\begin{align}
	f(\mathbf{r})
	=
	\frac{1}{\sqrt{N_c}}\sum_\mathbf{k} e^{i\mathbf{k}\cdot \mathbf{r}}f(\mathbf{k}),
\end{align}
the driving field in real space is simply
\begin{align}
	|f^l\rangle
	=
	\sum_{\mathbf{r}}
	f(\mathbf{r})
	\left(	
	l_A |A,\mathbf{r}\rangle
	+
	l_B |B,\mathbf{r}\rangle
	+
	l_C |C,\mathbf{r}\rangle
	\right).
\end{align}
Numerically and experimentally, this $|f^l\rangle$ is the driving field to be applied to the lattice.

Finally we comment on the detection scheme. Performing $q$ measurements with $q$ different driving fields $|f^l\rangle$, we want to measure the combined center of mass
\begin{align}
	\llangle \hat{r}_i \rrangle \equiv \frac{\sum_l \langle \beta^l | \hat{r}_i | \beta^l\rangle}{\sum_l \langle \beta^l | \beta^l\rangle}, \label{wanttomeasure}
\end{align}
where $|\beta^l\rangle$ is the steady-state under the driving field $|f^l\rangle$.
Expanding $|\beta^l\rangle$ as
\begin{align}
	|\beta^l\rangle
	=
	\sum_{\mathbf{r}}
	\left( \beta^l_{A,\mathbf{r}} |A,\mathbf{r}\rangle + \beta^l_{B,\mathbf{r}} |B,\mathbf{r}\rangle + \beta^l_{C,\mathbf{r}} |C,\mathbf{r}\rangle\right),
\end{align}
each term in the denominator of (\ref{wanttomeasure}) is
\begin{align}
	\langle \beta^l | \beta^l\rangle
	=
	\sum_\mathbf{r}
	\left(
	|\beta^l_{A,\mathbf{r}}|^2 + |\beta^l_{B,\mathbf{r}}|^2 + |\beta^l_{C,\mathbf{r}}|^2
	\right), \label{measuredenominator}
\end{align}
which is nothing but the total intensity of the fields summed over the whole lattice.
Each term in the numerator of (\ref{wanttomeasure}) is
\begin{align}
	\langle \beta^l |\hat{r}_i| \beta^l\rangle
	=
	\sum_\mathbf{r}
	r_i
	\left(
	|\beta^l_{A,\mathbf{r}}|^2 + |\beta^l_{B,\mathbf{r}}|^2 + |\beta^l_{C,\mathbf{r}}|^2
	\right), \label{measurenumerator}
\end{align}
which is the average $r_i$-position of the steady-state, up to a normalization factor.
Therefore $\llangle \hat{r}_i \rrangle$ can be numerically and experimentally measurable by detecting the intensity of each lattice site $|\beta^l_{A,\mathbf{r}}|^2$, $|\beta^l_{B,\mathbf{r}}|^2$, and $|\beta^l_{C,\mathbf{r}}|^2$. In the main text, when we numerically estimate the Berry curvature and the Fubini-Study metric for the 2D Lieb lattice, we use (\ref{measuredenominator}) and (\ref{measurenumerator}) to estimate $\llangle \hat{r}_i \rrangle$. The same expressions can be used to experimentally estimate $\llangle \hat{r}_i \rrangle$.
One important remark is that the factor $r_i$ in (\ref{measurenumerator}) is common to all the sublattices in the same unit cell. This means that we are not taking into account the relative positions of A, B, and C sublattices within a unit cell; upon estimating $\langle \beta^l |\hat{r}_i| \beta^l\rangle$, the position $r_i$ is the position assigned to each unit cell but not to each lattice site.

\section{II. Full derivation of the center of mass}

In the main text, we wrote down the expression of the combined center of mass $\llangle x_i \rrangle$ when the band is flat and contribution from other bands are negligible. We derive this result here, starting from a general situation when the band can have a non-flat dispersion and the other bands are present.

As written in the main text, the combined center of mass is
\begin{align}
	\llangle \hat{r}_i \rrangle
	=
	\frac{\sum_l \sum_\mathbf{k}\langle \beta^l(\mathbf{k})|i\partial_{k_i}|\beta^l(\mathbf{k})\rangle}{\sum_l \sum_\mathbf{k} \langle \beta^l(\mathbf{k})|\beta^l(\mathbf{k})\rangle}
	\equiv
	\frac{\llangle \hat{r}_i \rrangle_\mathrm{num}}{\llangle \hat{r}_i \rrangle_\mathrm{den}},
\end{align}
where the steady-state up to the first order in the external field is
\begin{align}
	&|\beta^l(\mathbf{k})\rangle
	=
	\frac{1}{(\omega_0 + i\gamma)\mathbb{I}_{q} - H_\mathbf{k}}
	|f^l(\mathbf{k})\rangle
	-i E_j
	\frac{1}{(\omega_0 + i\gamma)\mathbb{I}_{q} - H_\mathbf{k}}
	\frac{\partial}{\partial k_j}
	\frac{1}{(\omega_0 + i\gamma)\mathbb{I}_{q} - H_\mathbf{k}}
	|f^l(\mathbf{k})\rangle.
\end{align}
The derivative $\partial/\partial k_j$ acts on everything on its right.
Since Bloch states form a complete set of basis in the subspace of fixed $\mathbf{k}$, we have 
\begin{align}
	\mathbb{I}_q = \sum_n |n_\mathbf{k}\rangle \langle n_\mathbf{k}|,
\end{align}
where the sum is over all the bands.
Then, the steady-state can be written as
\begin{align}
	&|\beta^l(\mathbf{k})\rangle
	=
	\sum_n 
	\frac{|n_\mathbf{k}\rangle \langle n_\mathbf{k}|f^l(\mathbf{k})\rangle}{\omega_0 + i\gamma - \mathcal{E}_n(\mathbf{k})}
	- i E_j \sum_n
	\frac{1}{(\omega_0 + i\gamma)\mathbb{I}_{q} - H_\mathbf{k}}
	\frac{\partial}{\partial k_j}
	\frac{|n_\mathbf{k}\rangle \langle n_\mathbf{k}|f^l(\mathbf{k})\rangle}{\omega_0 + i\gamma - \mathcal{E}_n(\mathbf{k})}.
\end{align}
Using this expression, $\llangle \hat{r}_i \rrangle_\mathrm{den}$ becomes
\begin{align}
	\llangle \hat{r}_i \rrangle_\mathrm{den}
	&=
	\sum_l \sum_\mathbf{k} \langle \beta^l(\mathbf{k})|\beta^l(\mathbf{k})\rangle
	\notag \\
	&=
	\sum_{l,\mathbf{k},m,n}
	\left[
	\frac{\langle n_\mathbf{k}| \langle f^l(\mathbf{k}) |n_\mathbf{k}\rangle}{\omega_0 - i\gamma - \mathcal{E}_n(\mathbf{k})}
	+i E_j
	\frac{\partial}{\partial k_j}
	\left(
	\frac{\langle n_\mathbf{k}| \langle f^l(\mathbf{k}) | n_\mathbf{k}\rangle}{\omega_0 - i\gamma - \mathcal{E}_n(\mathbf{k})}
	\right)
	\frac{1}{(\omega_0 - i\gamma)\mathbb{I}_{q} - H_\mathbf{k}}
	\right]
	\notag \\
	&\hspace{2cm}\left[ 
	\frac{|m_\mathbf{k}\rangle \langle m_\mathbf{k}|f^l(\mathbf{k})\rangle}{\omega_0 + i\gamma - \mathcal{E}_m(\mathbf{k})}
	-i E_j
	\frac{1}{(\omega_0 + i\gamma)\mathbb{I}_{q} - H_\mathbf{k}}
	\frac{\partial}{\partial k_j}
	\left(
	\frac{|m_\mathbf{k}\rangle \langle m_\mathbf{k}|f^l(\mathbf{k})\rangle}{\omega_0 + i\gamma - \mathcal{E}_m(\mathbf{k})}
	\right)
	\right]
	\notag \\
	&=
	\sum_{l,\mathbf{k},n}
	\frac{\langle n_\mathbf{k}|f^l(\mathbf{k})\rangle \langle f^l(\mathbf{k}) |n_\mathbf{k}\rangle}{|\omega_0 + i\gamma - \mathcal{E}_n(\mathbf{k})|^2}
	\notag \\
	&+iE_j
	\sum_{l,\mathbf{k},m,n}
	\left\{
	\frac{\partial}{\partial k_j}
	\left(
	\frac{\langle n_\mathbf{k}| \langle f^l(\mathbf{k}) | n_\mathbf{k}\rangle}{\omega_0 - i\gamma - \mathcal{E}_n(\mathbf{k})}
	\right)
	\frac{|m_\mathbf{k}\rangle \langle m_\mathbf{k}|f^l(\mathbf{k})\rangle}{|\omega_0 + i\gamma - \mathcal{E}_m(\mathbf{k})|^2}
	-
	\frac{\langle n_\mathbf{k}| \langle f^l(\mathbf{k}) |n_\mathbf{k}\rangle}{|\omega_0 - i\gamma - \mathcal{E}_n(\mathbf{k})|^2}
	\frac{\partial}{\partial k_j}
	\left(
	\frac{|m_\mathbf{k}\rangle \langle m_\mathbf{k}|f^l(\mathbf{k})\rangle}{\omega_0 + i\gamma - \mathcal{E}_m(\mathbf{k})}
	\right)
	\right\}
	\notag \\
	&+\mathcal{O}(E_j^2).
\end{align}
Now we write $|f^l(\mathbf{k})\rangle = f(\mathbf{k})|l\rangle$ and use the completeness relation $\sum_{l=1}^q |l\rangle \langle l | = \mathbb{I}_q$. Up to the first order in $E_j$,
\begin{align}
	\llangle \hat{r}_i \rrangle_\mathrm{den}
	&=
	\sum_{\mathbf{k},n}
	\frac{f(\mathbf{k})^2}{|\omega_0 + i\gamma - \mathcal{E}_n(\mathbf{k})|^2}
	\notag \\
	&+iE_j
	\sum_{l,\mathbf{k},m,n}
	\left\{
	\frac{\partial}{\partial k_j}
	\left(
	\frac{\langle n_\mathbf{k}| \langle f^l(\mathbf{k}) | n_\mathbf{k}\rangle}{\omega_0 - i\gamma - \mathcal{E}_n(\mathbf{k})}
	\right)
	\frac{|m_\mathbf{k}\rangle \langle m_\mathbf{k}|f^l(\mathbf{k})\rangle}{|\omega_0 + i\gamma - \mathcal{E}_m(\mathbf{k})|^2}
	-
	\frac{\langle n_\mathbf{k}| \langle f^l(\mathbf{k}) |n_\mathbf{k}\rangle}{|\omega_0 - i\gamma - \mathcal{E}_n(\mathbf{k})|^2}
	\frac{\partial}{\partial k_j}
	\left(
	\frac{|m_\mathbf{k}\rangle \langle m_\mathbf{k}|f^l(\mathbf{k})\rangle}{\omega_0 + i\gamma - \mathcal{E}_m(\mathbf{k})}
	\right)
	\right\}
	\notag \\
	&=
	\sum_{\mathbf{k},n}
	\frac{f(\mathbf{k})^2}{|\omega_0 + i\gamma - \mathcal{E}_n(\mathbf{k})|^2}
	+iE_j
	\sum_{\mathbf{k},n}
	i\gamma
	\frac{\partial}{\partial k_j}
	\frac{f(\mathbf{k})^2}{|\omega_0 + i\gamma - \mathcal{E}_n(\mathbf{k})|^4}.
\end{align}
The last term is a total derivative term, and vanishes upon summing over $\mathbf{k}$. Therefore, we obtain
\begin{align}
	\llangle \hat{r}_i \rrangle_\mathrm{den}
	=
	\sum_{\mathbf{k},n}
	\frac{f(\mathbf{k})^2}{\left( \omega_0 - \mathcal{E}_n(\mathbf{k})\right)^2 + \gamma^2}
	+\mathcal{O}(E_j^2).
\end{align}
We can similarly obtain an expression for $\llangle \hat{r}_i \rrangle_\mathrm{num}$. However, the calculation is much more tedious, so we just write down the result here. Up to the first order in $E_j$, we have
\begin{align}
	\llangle \hat{r}_i \rrangle_\mathrm{num}
	=&
	\sum_{\mathbf{k},n} \frac{f(\mathbf{k})^2 \gamma \partial_{k_i} \mathcal{E}_n (\mathbf{k})}{[ \left( \omega_0 - \mathcal{E}_n(\mathbf{k})\right)^2 + \gamma^2 ]^2}
	-E_j
	\sum_{\mathbf{k},n}
	\frac{f(\mathbf{k})^2}{[ \left( \omega_0 - \mathcal{E}_n(\mathbf{k})\right)^2 + \gamma^2 ]^2}
	\left[
	\gamma \Omega_{ij}^n(\mathbf{k} + (\omega_0 - \mathcal{E}_n (\mathbf{k})) 2 g_{ij}^n (\mathbf{k})
	\right]
	\notag \\
	&- E_j \sum_{\mathbf{k},n}
	\left\{
	\frac{2 (\omega_0 - \mathcal{E}_n(\mathbf{k}))}{[ \left( \omega_0 - \mathcal{E}_n(\mathbf{k})\right)^2 + \gamma^2 ]^2}
	\partial_{k_i} f(\mathbf{k}) \partial_{k_j} f(\mathbf{k})
	+
	\frac{2 (\omega_0 - \mathcal{E}_n(\mathbf{k}))}{[ \left( \omega_0 - \mathcal{E}_n(\mathbf{k})\right)^2 + \gamma^2 ]^3}
	f(\mathbf{k})^2 \partial_{k_i} \mathcal{E}_n(\mathbf{k}) \partial_{k_j} \mathcal{E}_n (\mathbf{k})
	\right.
	\notag \\
	&\hspace{1cm}
	\left.
	+
	\frac{(\omega_0 - \mathcal{E}_n(\mathbf{k}))^2}{[ \left( \omega_0 - \mathcal{E}_n(\mathbf{k})\right)^2 + \gamma^2 ]^3}
	\left(
	\partial_{k_i} f(\mathbf{k})^2 \partial_{k_{j}} \mathcal{E}_n (\mathbf{k})
	+
	\partial_{k_j} f(\mathbf{k})^2 \partial_{k_{i}} \mathcal{E}_n (\mathbf{k})
	\right)
	\right.
	\notag \\
	&\hspace{1cm}
	\left.
	-
	\frac{\gamma^2}{[ \left( \omega_0 - \mathcal{E}_n(\mathbf{k})\right)^2 + \gamma^2
	 ]^3}
	\left(
	\partial_{k_i} f(\mathbf{k})^2 \partial_{k_{j}} \mathcal{E}_n (\mathbf{k})
	-
	\partial_{k_j} f(\mathbf{k})^2 \partial_{k_{i}} \mathcal{E}_n (\mathbf{k})
	\right)
	\right\}
	\notag \\
	&+ E_j \sum_{\mathbf{k}, n\neq m}
	f(\mathbf{k})^2
	\left\{
	\frac{\langle \partial_{k_i} n_\mathbf{k} | m_\mathbf{k}\rangle \langle m_\mathbf{k}|\partial_{k_j} n_\mathbf{k}\rangle}{(\omega + i\gamma - \mathcal{E}_m(\mathbf{k}))^2}
	\frac{1}{\omega_0 - i\gamma - \mathcal{E}_n (\mathbf{k})}
	+
	\frac{\langle \partial_{k_j} n_\mathbf{k} | m_\mathbf{k}\rangle \langle m_\mathbf{k}|\partial_{k_i} n_\mathbf{k}\rangle}{(\omega - i\gamma - \mathcal{E}_m(\mathbf{k}))^2}
	\frac{1}{\omega_0 + i\gamma - \mathcal{E}_n (\mathbf{k})}
	\right\}
	\notag \\
	&+\mathcal{O}(E_j^2).
\end{align}
This is a general expression which is always valid. Now we want to simplify it by assuming that the driving frequency $\omega_0$ is close to the energy of one band $\mathcal{E}_n$, which is sufficiently flat, and the other bands are sufficiently separated so that their contributions can be neglected.
Assumption of the band being flat sets all the group velocity terms (terms of the form $\partial_{k_i}\mathcal{E}_n (\mathbf{k})$) to zero. As in the main text, we define the detuning of the driving field from the band energy by $\Delta \equiv \omega_0 - \mathcal{E}_n$.
Then, we obtain
\begin{align}
	\llangle \hat{r}_i \rrangle_\mathrm{den}
	=
	\sum_{\mathbf{k}}
	\frac{f(\mathbf{k})^2}{\Delta^2 + \gamma^2}
	+\mathcal{O}(E_j^2)
\end{align}
and
\begin{align}
	\llangle \hat{r}_i \rrangle_\mathrm{num}
	=&
	-\frac{E_j}{[ \Delta^2 + \gamma^2 ]^2}
	\sum_{\mathbf{k}}
	\left\{
	f(\mathbf{k})^2
	\left[
	\gamma \Omega_{ij}^n(\mathbf{k}) + 2 \Delta g_{ij}^n (\mathbf{k})
	\right]
	+
	2 \Delta
	\partial_{k_i} f(\mathbf{k}) \partial_{k_j} f(\mathbf{k})
	\right\}
	+\mathcal{O}(E_j^2)
\end{align}
as in the main text.
The combined center-of-mass position is given by $\llangle \hat{r}_i \rrangle = \llangle \hat{r}_i \rrangle_\mathrm{num}/\llangle \hat{r}_i \rrangle_\mathrm{den}$.
Now we consider two specific driving schemes and simplify the expression.

\subsection{A. Single-site driving}

When the driving is single-site in real space, the momentum space driving profile $f(\mathbf{k})$ becomes independent of $\mathbf{k}$.
Then, up to the lowest order in $E_j$, 
\begin{align}
	\llangle \hat{r}_i\rrangle
	=
	-\frac{1}{\mathcal{A}}\frac{E_j}{\Delta^2 + \gamma^2}
	\left[
	\gamma\sum_\mathbf{k} \Omega^n_{ij} (\mathbf{k})
	+
	2\Delta \sum_\mathbf{k} g^n_{ij} (\mathbf{k})
	\right],
\end{align}
where $\mathcal{A}$ is the area of the Brillouin zone in momentum space.
Now, the steady-state response depends on the average of the Berry curvature, which is proportional to the Chern number, and the Fubini-Study metric. One can separate the contributions from the Berry curvature and the Fubini-Study metric by performing two sets of measurements interchanging $i$ and $j$, taking the sum or the difference, making use of $\Omega^n_{ij} = -\Omega^n_{ji}$ and $g^n_{ij} = g^n_{ji}$. This driving scheme was previously used to estimate the Chern number of driven-dissipative lattices~[S1-S3].

\subsection{B. Gaussian driving}

In the main text, we have considered the case where $f(\mathbf{k})$ is a Gaussian of the form 
\begin{align}
	f(\mathbf{k})
	=
	f \exp \left( -(\mathbf{k} - \mathbf{k}_0)^2 / 2\sigma_k^2\right).
\end{align}
Assuming that this Gaussian is sharply peaked around $\mathbf{k}_0$, the sum over $\mathbf{k}$ picks up the value of $\Omega_{ij}^n (\mathbf{k})$ and $g_{ij}^n (\mathbf{k})$ at $\mathbf{k}_0$. Namely, a part in $\llangle \hat{r}_i \rrangle_\mathrm{num}$ becomes simplified as
\begin{align}
	&\sum_{\mathbf{k}}
	f(\mathbf{k})^2
	\left[
	\gamma \Omega_{ij}^n(\mathbf{k}) + 2 \Delta g_{ij}^n (\mathbf{k})
	\right]
	\approx
	\left[
	\gamma \Omega_{ij}^n(\mathbf{k}_0) + 2 \Delta g_{ij}^n (\mathbf{k}_0)
	\right]
	\sum_{\mathbf{k}}
	f(\mathbf{k})^2
\end{align}
Since the derivative of $f(\mathbf{k})$ obeys $\partial_{k_i} f(\mathbf{k}) = -(k_i - k_{0i})f(\mathbf{k})/\sigma_k^2$, we obtain
\begin{align}
	\llangle \hat{r}_i \rrangle
	\approx&
	-\frac{E_j}{\Delta^2 + \gamma^2}
	\left[
	\gamma \Omega_{ij}^n(\mathbf{k}_0) + 2 \Delta g_{ij}^n (\mathbf{k}_0)
	\right]
	-\frac{E_j}{\Delta^2 + \gamma^2}\frac{2\Delta}{\sigma_k^4}
	\frac{\sum_\mathbf{k} (k_i - k_{0i}) (k_j - k_{0i}) f(\mathbf{k})^2}{\sum_\mathbf{k} f(\mathbf{k})^2}
	\notag \\
	=&
	-\frac{E_j}{\Delta^2 + \gamma^2}
	\left[
	\gamma \Omega_{ij}^n(\mathbf{k}_0) + 2 \Delta g_{ij}^n (\mathbf{k}_0)
	+
	\delta_{ij}\frac{\Delta}{\sigma_k^2}
	\right],
\end{align}
which is the result shown in the main text.

\end{widetext}
\end{document}